\documentclass[aps,prd,amsmath,amssymb,12pt]{revtex4}
\usepackage{bm}
\usepackage{tcolorbox}
\def\journal#1#2#3#4{{#1} {\bf #2}, #3 (#4)}
\newcommand{\be}{\begin{equation}}
\newcommand{\ee}{\end{equation}}
\newcommand{\bea}{\begin{eqnarray}}
\newcommand{\eea}{\end{eqnarray}}
\newcommand{\hf}{\frac12}
\newcommand{\nn}{\cr}
\def\eq#1{(\ref{#1})}
\def\la{\langle}
\def\ra{\rangle}
\def\Tr{{\mathrm{Tr}}}

\def\mr#1{{\mathrm{#1}}}

\def\ord#1{{\cal O}(#1)}
\def\ih{\frac{i}{\hbar}}

\def\hj{\tilde j}
\def\tx{\tilde x}
\def\hy{\tilde y}

\def\tx{\tilde x}

\begin{document}
\title{Equilibrium properties and decoherence of an open harmonic oscillator}
\author{Janos Polonyi}
\affiliation{Strasbourg University, CNRS-IPHC,23 rue du Loess, BP28 67037 Strasbourg Cedex 2 France}
\begin{abstract}
The equilibrium properties of an open harmonic oscillator are considered in three steps: First the creation and destruction operators are generalized for open dynamics and the creation operator is used to construct coherent states. The second step consists of the introduction of the Heisenberg representation where the dynamical decoherence is identified. Finally it is pointed out that the quantum fluctuations generate non-continuous limit for infinitesimal system-environment interactions and at the border of the under- and over-damped oscillator.
\end{abstract}
\date{\today}
\maketitle

\section{Introduction}
The systematic construction of the quantization rules for open system is based on the elimination of the environmental degrees of freedom. However this is a quite involved procedure therefore it is important to understand better some simple, generic examples. The simplest open dynamics corresponds to weak interactions within the system and between the system and its environment. The leading approximation to such a dynamics is a harmonic effective theory and the goal of this work is to gain some insight into the relaxed dynamics of an weakly open harmonic oscillator. 

The problem of an open harmonic oscillator has attracted lot of attention and impressing advances have already been made. After the initial steps, using the quantum Fokker-Planck \cite {agarwal} and Langevin equations \cite{ford,gardinerc,dattagupta} a simplified master equation was put forward \cite{walls,savage}, to be generalized to a non-local master equation by the elimination of the environment \cite{haake,caldeira,riseborough}. The following large number of works, using the operator and the path integral formalism is reviewed in refs. \cite{dekker} and \cite{grabert}, respectively. A specially difficult problem is to treat the non-local nature in time of an effective dynamics. The non-Markovian master equation for the reduced density matrix can be found by the help of the projector operator method \cite{nakajima, zwanzig}. The technical difficulties can partly be reduced by introducing correlated system-environment states \cite{esposito,budini,breuerpr} and the collisional picture \cite{vacchini}. In the case of the harmonic oscillator the non-locality in time leads to the absence of unique local master equation \cite{karrlein}. The general theory of open systems has been presented in excellent textbooks \cite{gardiner,breuer}. An accessible formal treatment of quantum semigroups is given in \cite{alicki}. A simple, readable introduction with a number of interesting application is \cite{banerjee}.

We rely in this work on the effective theory for the oscillator within the path integral formalism which offers new approximation schemes: The non-locality in time can be represented by retaining either the higher order terms in the time derivatives \cite{boyankovsky} or true multi-local terms \cite{alf} in the effective equation of motion. The former scheme leads to instabilities owing to Ostrogadsky's theorem \cite{ostrogadsky} and the demonstration of stability remains a numerical issue in the latter. We are satsfied by retaining the first two order of the derivative expansion thereby Ostrogadsky's instability is avoided. The leading order term represents Newton's friction force, hence our scheme can be considered as a consistent generalization of the friction force as a phenomenological tool of classical mechanics, to open quantum systems.

The closed harmonic oscillator has a discrete, equidistant spectrum and the simplest way to explore its stationary states is to use the creation and destruction operators of the elementary excitations. The main goal of the present paper is find the generalization of the stationary states, the creation and destruction operators and the Heisenberg representation for an open harmonic oscillator. These operators have already been considered in previous works. The creation and the annihilation operators for the solution of the master equation of a harmonic oscillator, linearly coupled to other oscillators with Ohmic spectral function has been constructed by the help of action-angle variables \cite{tay}. The diagonalization of the generator of the time evolution for a system of $n$ fermions, following a harmonic open dynamics \cite{prosen} has been used to define the creation and annihilation operators. The ladder operators of the restricted master equation, obtained in the van Hove limit \cite{honda} has been found, as well. A more phenomenological approach is followed in this work where our goal is to find the ladder operators for the most general open harmonic oscillator with local, Markovian dynamics. This strategy is realized within the framework of the Closed Time Path (CTP) formalism \cite{schw,keldysh} which is well suited to our goal since it offers a natural way to find the reduced system density matrix. The full, exact master equation has already been derived in this formalism for a particle, coupled linearly to a set of harmonic oscillators \cite{hu9293}. The  non-local memory in time of the exact  master equation renders the use of the result difficult. A more phenomenological starting point, leading to a simpler however more useful scheme, is to consider the most general harmonic local, i.e. Markovian open dynamics in time. This dynamics is defined in this work by an effective Lagrangian. The Fokker-Planck equation, corresponding to this Lagrangian, agrees with the most general master equation for harmonic systems \cite{sandulescu,isar}. The additional insight, obtained from such a derivation of the master equation, is that there are total time derivative terms in the Lagrangian with kinematic rather than dynamical role on the master equation level. A further bonus of the CTP formalism is a clear and natural way the decoherence can be accessed \cite{dyndec}. 

We use the most general harmonic, Markovian master equation in this work in a manner as close as possible to Schr\"odinger's equation of a closed harmonic oscillator. While the bra and the ket dynamics are independent in a closed systems it is enough to follow one of them because they are complex conjugate of each other. However the bra and ket components are coupled by the non-separable terms of the density matrix of a mixed state hence they must be handled separately, independently of each other. This is the physical source of the apparent reduplication of the  degrees of freedom in the CTP formalism. The density matrix of a harmonic system, a Gaussian in the two coordinate variables, is reminiscent of the wave function of two harmonic oscillators. This similarity motivates to work in the Liouville space of operators and to construct the ladder operators, acting on the space of density matrices and shifting the eigenvalue of the generator of the time evolution. In the case of a closed oscillator this procedure yields the traditional creation and destruction operators which shift the eigenvalue by $\pm i$ times the energy of an elementary excitation and handle pure states by acting exclusively either on the bra or on the ket. The main result of this work is a generalization of the ladder operators for open oscillator. They generate complex shifts of the eigenvalue, the real part representing dissipative forces and the excitations, created or destroyed, correspond to mixed density matrix, in other words they correlate the quantum fluctuations in the bra and the ket. One arrives in this manner at a rather simple diagonal form of the generator of the time evolution, a suggestive generalization of the case of the closed oscillator. The relaxed state turns out to be unique within the Hilbert space of states, defined by the given asymptotic conditions at spatial infinity. The coherent states are constructed as a simple application of this scheme. 

The Heisenberg representation is usually defined on the level of the full, closed system, following conservative, local dynamics. It would be desirable to construct this representation within the much smaller linear space f the observed system. This is not an obvious procedure owing to the non-unitary effective system dynamics. Another result of this work is to show that the conservation of the total probability is sufficient to find a simple extension of the Heisenberg representation for an arbitrary master equation, either local or non-local in time.

The decoherence stands for the suppression of the different interference terms in the expectation values. The usual signature, the decay of the density matrix elements with increasing off-diagonality \cite{zeh,zurekd,joos}, characterizes the instantaneous decoherence of the actual state and can easiest be seen in the Schr\"odinger representation. Another result, reported below is a simple and natural identification of the dynamical decoherence in the Heisenberg representation which is closer to our intuitive ideas about the environment induced decoherence.

The Markovian master equations are linear and their stationary eigenvectors are defined by their particular time-dependence of the density matrix, namely an exponential prefactor. While these eigenvectors are not physical states except the relaxed, time-independent eigenvector they represent a basis set to construct the physical states \cite{free}. The construction of the stationary  eigenvectors reveals a genuine quantum effect of the open dynamics, related to the degeneracy of the spectrum. The classical dynamics of a damped harmonic oscillator and as a result the first moments of the canonical operators are analytic in Newton's friction coefficient, the only classical environmental parameter. However the higher moments of the canonical operators contain the quantum fluctuations and display singularities in the environment parameters. These singularities appear at infinitesimal system-environment interactions and at the border between the under- and over-dumped oscillator.

We start with a schematic derivation of the harmonic master equation in section \ref{mastereqs}. The relaxed state and a simple set of creation and destruction operators are defined in section \ref{nonstatb}. The ladder operators which create and destruct stationary excitations are the subject of section \ref{statsts}, followed by the definition of the coherent state in section \ref{cohsts}. The extension of the Heisenberg representation for open systems is presented in section \ref{heisenbs} and is used in section \ref{decs} to gain a fresh view on decoherence. The singularities, generated by the quantum fluctuations at the degeneracy of the master equation are mentioned in section \ref{sings}. Finally section \ref{sums} contains the summary of our results.

\section{Master equation}\label{mastereqs}
The master equation for the reduced density matrix of the observed system can be obtained in two steps, by constructing the effective Lagrangian followed by the derivation of the equation of motion, corresponding to it.

\subsection{Effective Lagrangian}
The effective theory for the reduced density matrix is the easiest to derive in the CTP path integral formalism. Let us suppose that the observed system and its environment together make up a closed dynamics, defined by the action $S[x,y]$ in the path integral expressions where $x$ and $y$ denote the system and the environment coordinates, respectively. The full density matrix, 
\be\label{fulldm}
\rho(t)=U(t,t_i)\rho(t_i)U^\dagger(t,t_i),
\ee
given in terms of the time evolution operator $U(t,t_i)$, can obtained by integrating over the trajectory pairs, $\tx=(x_+,x_-)$, $\hy=(y_+,y_-)$,
\be
\la x_+,y_+|\rho(t)|x_-,y_-\ra=\int D[\tx]D[\hy]e^{\ih S[x_+,y_+]-\ih S[x_-,y_-]}\la x_{i,+},y_{i,+}|\rho(t_i)|x_{i,-},y_{i,-}\ra
\ee
where $x_\pm(t)=x_\pm$, $y_\pm(t)=y_\pm$ and $x_\pm(0)=x_{i,\pm}$, $y_\pm(0)=y_{i,\pm}$. The reduced system density matrix,
\be\label{densmte}
\rho(x_+,x_-)=\la x_+|\Tr_y[U(t,t_i)\rho_iU^\dagger(t,t_i)]|x_-\ra,
\ee
can be written in the form
\be
\rho(x_+,x_-)=\int D[\tx]D[\hy]e^{\ih S[x_+,y_+]-\ih S[x_-,y_-]}
\ee
where the trajectory $\hy$ is closed, $y_+(t)=y_-(t)$ and the convolution with the initial density matrix, assumed to be factorisable in the system and its environment, is suppressed. By separating the system and the environment action, $S[x,y]=S_s[x]+S_e[x,y]$, one defines the effective system action, $S_{eff}[\tx]=S_s[x_+]-S_s[x_-]+S_{infl}[\tx]$, which characterizes the system dynamics,
\be\label{reddensm}
\rho(x_+,x_-)=\int D[\tx]e^{\ih S_{eff}[\tx]},
\ee
where the influence functional is given by 
\be
e^{\ih S_{infl}[\tx]}=\int D[\hy]e^{\ih S_e[x_+,y_+]-\ih S_e[x_-,y_-]}.
\ee

The unitarity of the time evolution of the observed system plus its environment preserves of the total probability, $\Tr[\rho]=1$, keeps $\rho$ Hermitean and implies the CTP symmetry, $S_{infl}[x_-,x_+]=-S^*[x_+,x_-]$. Another important consequence of the relation $UU^\dagger=\openone$ is that the final time can be sent to infinity and the expectation value of are obtained by the help of the generator functional,
\be\label{genfunc}
Z[\hj]=\int D[\tx]e^{\ih S_{eff}[\tx]+\ih\int dt\hj\tx},
\ee
where the integration is over the CTP trajectory pairs, spanning  $-\infty<t<\infty$. It proves to be advantageous to use the coordinates $x=(x_++x_-)/2$ and $x_d=x_+-x_-$, representing the average physical coordinate and the quantum fluctuations, respectively.

We are interested in local harmonic effective dynamics where the influence functional arises from a local influence Lagrangian. The latter can be obtained in a phenomenological manner by the help of the Landau-Ginzburg double expansion, using the coordinate (measured from the equilibrium position) and the inverse of its characteristic time, or time derivative, as small parameters. The Lagrangian is truncated at the second order in these small parameters. Since the higher order derivatives are excluded due to the instability they generate \cite{ostrogadsky} the dissipative effects of the environment are summarized in Newton's friction constant in this approximation. The most general quadratic Lagrangian consists of the terms $\dot x\dot x_d$, $\dot x^2$, $\dot x^2_d$, $x\dot x$, $x\dot x_d$, $x_d\dot x$, $x_d\dot x_d$, $x^2$, $x_dx$ and $x^2_d$. The CTP symmetry excludes $\dot x^2$, $x\dot x$ and $x^2$. The coefficient of the time derivatives are given in the mid-point prescription \cite{gas}. The total time derivatives lead to boundary contributions to the action \cite{schulman} and influence the expectation values in a time-independent manner. There are two total time derivative terms, $-\alpha(x\dot x_d+x_d\dot x)-i\beta x_d\dot x_d$, they generate the transformation
\be\label{gaugetr}
\rho\to e^{-i\frac\alpha{2\hbar}(x^2_+-x^2_-)+\frac\beta{2\hbar}(x_+-x_-)^2}\rho.
\ee
The effect of the first term in the exponent is a gauge transformation which can be taken into account by changing the momentum operator, $p\to p+\alpha x$, in the observables. We set $\alpha=0$ below for the sake of simplicity. The $\beta$-dependent term is not a gauge transformation and is retained, it introduces a static decoherence ($\beta<0$) or recoherence ($\beta>0$) in the coordinate basis. Our Lagrangian therefore is chosen to be
\be\label{efflagr}
L=m\left[\dot x\dot x_d-\omega^2xx_d-\nu\dot xx_d+\frac{i}2(d_0x^2_d+d_2\dot x^2_d)-i\beta x_d\dot x_d\right].
\ee
The equation of motion for $x$, obtained by varying $x_d$, indicates that $m\nu$ is Newton's friction constant. The parameters $d_0$ and $d_2$ control the decoherence in the coordinate basis and have to be non-negative to render the path integral convergent. This phenomenologically justified Lagrangian, apart of the total time derivative term, can be derived in two models \cite{gas}. In a harmonic toy model \cite{caldeira} the environment consists of infinitely many harmonic oscillators, coupled linearly to the system coordinate and characterized by a particular, Drude spectral function. In another, more realistic model a test particle interacts with an ideal gas and this Lagrangian is reproduced in the leading order of the particle-gas interaction and the $\ord{\partial_t^2}$ order of the expansion in the time derivative. The environment parameters, $\nu,d_0$ and $d_2$ turn out to be proportional to the square of the system-environment coupling strength, $g$, i.e. $\nu,d_0,d_2=\ord{g^2}$. Furthermore they are given by different expressions of the model parameters hence are considered below as independent phenomenological parameters.

\subsection{Equation of motion}
The equation of motion for the reduced density matrix can be derived by making an infinitesimal time step, $t\to t+\Delta t$, in the path integral \eq{reddensm}. The resulting ``Fokker-Planck equation'',
\be\label{master}
\partial_t\rho={\cal L}\rho,
\ee
contains the generator of the time evolution,
\be\label{genmast}
{\cal L}=i\nabla\nabla_d-ixx_d-\frac{d_0+d_2\nu^2-2\nu\beta}2x^{d2}+(d_2\nu-\beta)ix_d\nabla-\nu x_d\nabla_d+\frac{d_2}2\Delta,
\ee
written in terms of dimensionless variables $t\to t/\omega$, $x\to x\ell_{cl}$ where $\ell_{cl}=\sqrt{\hbar/m\omega}$, $\nu\to\omega\nu$, $d_0\to\omega^2d_0$, $d_2\to d_2$, $\beta\to\omega\beta$ and the derivatives $\nabla=\partial/\partial x$ and $\nabla_d=\partial/\partial x_d$. 

The justification of a phenomenological approach, leading to this generator, is far more involved as in the case of a closed system. Any Hamiltonian in Schr\"odinger's equation which is Hermitian and bounded from below is acceptable since it represents a possible physical system. There are few obvious necessary conditions on the generator of the time evolution for open system, ${\cal L}$, however the sufficient conditions are far from being clear. It is obvious that the generator must preserve the trace and the positivity of the density matrix. The difficulties about the sufficient conditions arise from the interaction of the observed system with an unobserved environment. This interaction makes the effective equation of motion for the observed system non-local in time. The usual remedy to avoid such a complication is to fall back on the Markovian approximation and ignore the non-local features in time in the effective dynamics. This amounts to the reduction of the exact integro-differential equation to a first order differential equation. However such an equation possesses a freely adjustable initial condition and additional considerations are needed to find the range of initial conditions which is compatible with our truncation of the effective dynamics. The reduction of the non-local features in time to a differential equation is the strategy of the expansion in the derivative and is expected to be reasonable for slow enough time dependence in the system with respect to its environment. Assuming that this condition is met the allowed initial condition set should contain the nonentangled, factorizable system-environment states. 

Instead of searching of sufficient conditions for the master equation one can restrict the phenomenologically relevant equations by finding further necessary conditions. One possibility is to try to extend the allowed initial conditions over certain entangled initial states. For this to make sense one needs a condition to assure that the given effective equation of motion can be extended to a closed dynamics in a larger Hilbert space. One would think that this issue goes beyond the structure of the truncated, local effective equation of motion for the observed system. But there is actually a condition on the master equation, the complete positivity, which indicates the possibility of finding such an extension. The time dependence $\rho(t)$ of the reduced density matrix satisfies the complete positivity condition \cite{stienspring,kraus} if there is a set of time-dependent operators, $\{W_n(t)\}$, satisfying
\be
\openone=\sum_nW_n(t)W_n^\dagger(t)
\ee
and reproducing the given time dependence,
\be
\rho(t)=\sum_nW_n(t)\rho(0)W_n^\dagger(t).
\ee
The linear space, associated to the quantum number $n$ provides a minimal representation of an effective environment which generates the necessary system-environment entanglement. If the complete positivity condition is met then the effective system dynamics can be extended over the linear space of the index $n$ to a unitary closed dynamics. While the set of allowed initial condition now contains non-entangled states the extent of the allowed entangled initial states remains an open question. The completely positive, trace preserving master equations can be characterized by the Linblad structure \cite{gorini,lindblad}. The generator \eq{genmast} is the most general harmonic expression with this structure \cite{sandulescu,isar} as long as $\nu^2\le2d_0d_2$ apart of a time-independent gauge transformation which is quadratic in the coordinate.

A more natural and simpler way to justify the form \eq{genmast} of the generator is its microscopic derivation which shows clearly the physical origin. However the argument is incomplete owing to the truncations, carried out in the derivation. This problem can be avoided by relying on systematic truncations where the orders of a small parameter are either fully retained or fully omitted. In fact, any condition, expressing the existence of an environment is independent of the small parameter hence its systematically truncated form satisfies the same condition. The effective Lagrangian \eq{efflagr} is obtained by such an expansion hence the conditions on the expectation values of the test particle, derived in the full, closed system, remain valid on the truncated effective level. The environment parameters are given by different loop integrals, involving the test particle-gas interaction potential \cite{gas}. The free choice of the latter makes the environment parameters free phenomenological parameters of the master equation.

Our goal is to find the stationary eigenvectors of the master equation, defined by the eigenvalue condition ${\cal L}\rho_\Omega=\Omega\rho_\Omega$. Owing to the obvioius similarity with the stationary state condition for closed systems it is useful to employ the bra-ket notation in the Liouville space of operators \cite{masterl,mastertf} by considering the density matrix as a ``wave function'', $\rho(x,x_d)=\la\la x,x_d|\rho\ra\ra$, $\la\la x,x_d|\rho^\dagger\ra\ra=\rho^*(x,-x_d)$ and using the scalar product
\be\label{expval}
\la\la A|B\ra\ra=\Tr[\hat A^\dagger\hat B]=\int d\tx A^*(x,x_d)B(x,x_d).
\ee
To render the transition between the Liouville space operators and the usual operators, acting on the Hilbert space of pure states, we place a hat above the latter. The trace is associated to a particular bra, $\Tr[\hat A]=\la\la\Tr|A\ra\ra$, with $\la\la\Tr|x,x_d\ra\ra=\delta(x_d)$. Important properties of the trace are $\la\la\Tr|x_dO\ra\ra=\Tr[\hat x_d\hat O]=0$ and $\la\la\Tr|pO\ra\ra=\Tr[\hat p\hat O]=0$, where the operator $O$ is arbitrary and is restricted to have the same matrix elements for $x\to\pm\infty$, respectively. While the operators in the transition amplitudes of the closed dynamics are constructed by the help of the canonical pair, $(\hat x,\hat p)$, the scalar product in the space of operators needs two canonical pairs, $(x_\pm,p_\pm)$, namely $x_\pm\leftrightarrow\hat x$, $p_\pm=\mp i\nabla_\pm=\pm p/2+p_d\leftrightarrow\hat p$ where $p=-i\nabla$, $p_d=-i\nabla_d$ and the left and the right hand sides of the arrow display the Liouville and the Hilbert space expressions, respectively. Similar relations, needed below are $x\rho\leftrightarrow\{\hat x,\hat\rho\}/2$, $x_d\rho\leftrightarrow[\hat x,\hat\rho]$, $p\rho\leftrightarrow[\hat p,\hat\rho]$, $p_d\rho\leftrightarrow\{\hat p,\hat\rho\}/2$. The simplified expression for the momentum $p_\pm\to p_d\leftrightarrow\hat p$ can be used for normalizable states.

We need four different bases in the subspace of canonical operators, span by $x_\pm$ and $p_\pm$. The basis $X=(x,p,x_d,p_d)$ is useful in dealing with the solution of the master equation. The operators $c_\pm=(x_\pm+\nabla_\pm)/\sqrt2$ and $\bar c_\pm=(x_\pm-\nabla_\pm)/\sqrt2$ of the basis $C=(c_+,c_-,\bar c_+,\bar c_-)$ handle the elementary excitations of the closed dynamics and their representation in the Hilbert space is $c_+\rho\leftrightarrow\hat c\hat\rho$, $c_-\rho\leftrightarrow\hat\rho\hat c^\dagger$, $\bar c_+\rho\leftrightarrow\hat c^\dagger\hat\rho$ and $\bar c_-\rho\leftrightarrow\hat\rho\hat c$, with $\hat c=(\hat x+i\hat p)/\sqrt2$. The dagger is used for the operators, acting on the Hilbert space of pure states. Two further bases are introduced later.

Though the master equation is linear there is an essential difference in the use of the linear algebra in the space of pure states and density matrices.  Namely, the expectation values are linear in the density matrix, $\Tr[A\rho]=\la\la A^\dagger|\rho\ra\ra$, hence they do not contain the interference terms \cite{free}, $\la\la A^\dagger|(\rho_1+\rho_2)\ra\ra=\la\la A^\dagger|\rho_1\ra\ra+\la\la A^\dagger|\rho_2\ra\ra$. The basis where the density matrix is diagonal can be called decohered basis because the additive terms of the density matrix are completely decohered.

\section{Non-stationary basis}\label{nonstatb}
The general relaxed stationary solutions of the master equation with $\Omega=0$ can be specified by prescribing $\rho(x,0)$, $\rho(0,x_d)$ and $\nabla\rho(0,x_d)$ for $x_d\ge0$. We seek a solution in the linear space with a Gaussian asymptotic as $|x_\pm|\to\infty$, in particularly a simple Gaussian, 
\be\label{statdm}
\rho_0(x,x_d)=\frac{Q}{\sqrt{2\pi}}e^{-\frac{Q^2}2x^2-\frac{R^2}2x^2_d-iS^2x_dx}.
\ee
It is easy to see that the choice
\be\label{stdmpar}
Q^2=\frac{2\nu}{d_0+d_2},~~~
R^2=\frac{d_0+d_2}{2\nu}+\frac{2d_0d_2\nu}{d_0+d_2}-\beta,~~~
S^2=\frac{2d_2\nu}{d_0+d_2}
\ee
yields the desired solution. The positivity of the second moment of the canonical operators, $x_\pm$ and $p_\pm$,
\bea\label{x2p2}
\Tr[x_\pm^2\rho_0]&=&\frac{d_0+d_2}{2\nu},\nn
\Tr[p_\pm^2\rho_0]&=&\frac{d_0+d_2(1+\nu^2)}{2\nu}-\beta,
\eea
restricts the allowed range of the environmental parameters. The positivity of $\la x^2\ra$ represents no restricton as long as the path integral is convergent, $d_0,d_2\ge0$, the bound $\la p^2\ra>0$ defines the family of relaxed states each defining a linear space of excitations with Hermitean momentum operator. The Gaussian relaxed state supports the localisation length $\ell_{loc}=\sqrt{(d_0+d_2)/2\nu}$. The term with $S^2$ performs a gauge transformation and the density matrix is factorisable and corresponds to the ground state of a closed harmonic oscillator if $Q=2R$. 

It is advantageous to introduce the basis $B=(b,b_d,\bar b,\bar b_d)$ for the canonical operators where
\bea\label{bbd}
b&=&\frac{Qx+\frac{i}Q(p+S^2x_d)}{\sqrt2}\nn
b_d&=&i\frac{Rx_d+\frac{i}R(p_d+S^2x)}{\sqrt2},\nn
\bar b&=&\frac{Qx-\frac{i}Q(p+S^2x_d)}{\sqrt2},\nn
\bar b_d&=&-i\frac{Rx_d-\frac{i}R(p_d+S^2x)}{\sqrt2}.
\eea
because $b$ and $b_d$ annihilate the relaxed state, $b\rho_0=b_d\rho_0=0$. By the help of the creation operators one can build orthogonal vectors,
\be\label{cdest}
\rho^{(b)}_{m,n}=\frac{\bar b^m\bar b_d^n}{\sqrt{m!n!}}\rho_0
\ee
which are Hermitian owing to the Hermiticity of the relaxed state,
\be
\rho^*_{m,n}(x,-x_d)
=\frac{\bar b^m\bar b_d^n}{\sqrt{m!n!}}\rho_0^*(x,-x_d)
=\rho_{m,n}(x,x_d).
\ee
The linear space of the density matrices of the open oscillator is span by the vectors \eq{cdest}. This is not a specially useful basis set owing to its time-dependence, its importance is to assure that the stationary density matrix components, introduced at eq. \eq{diagest} below, form a basis, too.

\section{Stationary basis}\label{statsts}
The operator set $B$ is useful to generate a basis for the density matrix however the basis vectors follow a rather involved time-dependence. The stationary states can be obtained from $\rho_0$ by the help of the ladder operator set, $A$. These operators are defined by the commutation relation 
\be\label{ladder}
[{\cal L},A_\lambda]=\lambda A_\lambda,
\ee
and perform the shift, $\Omega\to\Omega+\lambda$, when acting on a stationary state. The density matrix of a harmonic system, being the product of bra and ket components, involves the normal frequencies of the classical dynamics forward and backward in time. Hence, the spectrum of \eq{ladder},
\be\label{spectrum}
\lambda_{\tau,\tau'}=-\tau\frac\nu2+\tau'i\omega_\nu,
\ee
where $\omega_\nu=\sqrt{1-\nu^2/4}$ and $\tau,\tau'=\pm1$, is on the unit circle on the complex frequency plan for an under-damped oscillator, $\nu<2$, and on the real axis in the over-damped case, $\nu>2$. It is worthwhile noting that the friction always generates a real part to the eigen frequencies hence the density matrix has no limit cycle.

The ladder operators, $A=(a_+,a_-,\bar a_+,\bar a_-)$, corresponding to the frequency shift $\lambda_{+,+},\lambda_{+,-},\lambda_{--+},\lambda_{-,+}$ listed in the same order, are 
\bea\label{apm}
a_\pm&=&N_\pm\left[\lambda_{+,\mp}x+i\frac{d_0\lambda_{+,\mp}+d_2\lambda_{-,\mp}}{2\nu}p+i\left(\beta-\frac{d_0+d_2}{2\nu}+\frac{d_2}2\lambda_{+,\pm}\right)x_d+p_d\right]\nn
\bar a_\pm&=&iN_\pm(\lambda_{+,\mp}p+x_d),
\eea
respectively, where $N_+=z$, $N_-=z^*$ with
\be\label{zfact}
z=\sqrt{\hf+\frac\nu{4i\omega_\nu}}.
\ee
These operators satisfy the commutation relations
\be\label{comm}
[a_\tau,\bar a_{\tau'}]=\delta_{\tau,\tau'}
\ee
and the equation $a_\tau\rho_0=0$ suggesting to interpret $a_\pm$ and $\bar a_\pm$ as annihilation and creation operators of the elementary excitations, respectively. A Gaussian density matrix \eq{statdm}, constructed by the help of different parameters than \eq{stdmpar} is not annihilated by $a_\pm$ and the Liouville space vector, obtained by acting on it with sufficiently many $a_\pm$ operators, contains components with exponentially increasing coefficient in time, a reminiscent of the unboundedness of the energy of a closed harmonic oscillator in a Hilbert space, constructed by acting the creation and annihilation operators on a state, different than the true ground state. An important identity, used repeatedly below, is $\Tr[\bar a_\pm O]=0$, it can be proven by partial integration and is valid for any operator $O$ with the same matrix elements as $x\to\pm\infty$.

We write the linear transformations, connecting the operator bases $X$, $A$, $B$ and $C$ in the form
\be
o=\sum_{o'}T_{o,o'}o'.
\ee
These transformations have unit determinant, $\det T=1$, we record here only few matrix elements,
\bea\label{atox}
T_{x,a_+}&=&T_{x,a_-}^*=\frac{i}{2z\omega_\nu},\nn
T_{x,\bar  a_+}&=&T_{x,\bar a_-}^*=z\frac{d_0\lambda_{+,-}+d_2\lambda_{-,-}}{2\nu},\nn
T_{p,a_\pm}&=&0,\nn
T_{p,\bar a_+}&=&-T_{p,\bar a_-}^*=\frac1{2z\omega_\nu},\nn
T_{x_d,a_\pm}&=&0,\nn
T_{x_d,\bar a_+}&=&-T_{x_d,\bar a_-}^*
=-iz,\nn
T_{p_d,a_+}&=&T_{p_d,a_-}^*
=z,\nn
T_{p_d,\bar a_+}&=&T_{p_d,\bar a_-}^*=z\left[\frac{d_0+d_2}{2\nu}+\frac{d_2}2\lambda_{+,+}-\beta\right],
\eea
and 
\bea\label{abtr}
T_{\bar a_+,\bar b}&=&T_{\bar a_-,\bar b}^*
=z\sqrt{\frac\nu{d_0+d_2}}\lambda_{-,+},\nn
T_{\bar a_+,\bar b_d}&=&T_{\bar a_-,\bar b_d}^*\nn
&=&-z\sqrt{\frac\nu{(d_0+d_2)[(d_0+d_2)^2-(d_0+d_2)2\beta\nu+d_0d_2\nu^2]}}[d_0+d_2(1+\nu\lambda_{-,+})],
\eea
used below.

The action of the ladder operators on the density matrix can be seen clearer by presenting the action of the ladder operators in the Hilbert space of pure states, $\bar a_\pm\rho\leftrightarrow[\hat{\bar a}_\pm,\hat\rho]$ and $a_\pm\rho\leftrightarrow[\hat a_\pm,\hat\rho]+\{\hat a'_\pm,\hat\rho\}$ with
\bea
\hat{\bar a}_\pm&=&\sqrt{\hf\pm\frac\nu{4i\omega_\nu}}[\pm(\omega_\nu\pm i\frac\nu2)p+x],\nn
\hat a_\pm&=&iN_\pm\left[\frac{d_0\lambda_{+,\mp}+d_2\lambda_{-,\mp}}{2\nu}p+\left(\beta-\frac{d_0+d_2(1+\nu\lambda_{+,\pm})}{2\nu}\right)x\right],\nn
\hat a'_\pm&=&\frac{N_\pm}2(\lambda_{+,\pm}x+p).
\eea
The ladder operators, written in the Hilbert space, are linear superpositions of the creation and annihilation operators of the closed dynamics, $\hat a_\pm$ and $\hat a_\pm^\dagger$, defined by the normal frequency $\lambda_{\mp,-}$, and correspond to mixed states with damping. 

The generator of the time evolution assumes a simple form,
\be\label{generator}
{\cal L}=\lambda_{-,-}\bar a_+a_++\lambda_{-,+}\bar a_-a_-,
\ee
in terms of the ladder operators with the eigenvector
\be\label{diagest}
\rho^{(a)}_{m,n}=\frac{\bar a_+^m\bar a_-^n}{\sqrt{m!n!}}\rho_0.
\ee
The Hermitian conjugate of the density matrix \eq{diagest} is $\rho^{\dagger(a)}_{m,n}=\rho^{(a)}_{n,m}$ and $\rho^{(a)}_{m,n}$ for $\nu<2$ and $\nu>2$, respectively. Hence the physical states are the linear superpositions of $\rho^{(a)}_{n,n}$, $\rho^{(a)}_{m,n}+\rho^{(a)}_{n,m}$ and  $i(\rho^{(a)}_{m,n}-\rho^{(a)}_{n,m})$ for the under-dumped and $\rho^{(a)}_{m,n}$ for the over-damped case, respectively. The density matrix $(4-\nu^2)^{n/2}\rho_{n,n}$ is positive since the operators $\Delta$, $ix_d\nabla$ and $-x_d^2$ have positive coefficients in the numerator of
\be
\bar a_+\bar a_-=\frac{\nabla^2+\nu ix_d\nabla-x^2_d}{\sqrt{4-\nu^2}},
\ee
as in the master equation which preserves the positivity. 

The generator of the time evolution, $\cal L$, has non-vanishing Hermitian and anti-Hermitian parts hence its eigenstates are not orthogonal and may not form a complete set. However the bases $A$ and $B$ are related by a non-singular liner transformation hence the  Liouville space vectors \eq{diagest} generate the same linear space as the one span by the basis vectors \eq{cdest}. In particular, the relaxed state with $\Omega=0$ is unique within the space of density matrices with the same asymptotic in the limit $|x|,|x_d|\to\infty$.

The linear superposition of the Liouville space vectors \eq{diagest} follows the time dependence
\be
\rho=\sum_{m,n}c_{m,n}e^{-[\frac\nu2(m+n)+i\omega_\nu(m-n)]t}\bar a_+^m\bar a_-^n\rho_0
\ee
The diagonal contributions, $m=n$, are suppressed by $e^{-m\nu t}$ as expected. The contributions of non-relaxed Liouville eigenstates do not contribute to the total probability since they have vanishing trace,
\be\label{traecons}
\Tr[\rho^{(a)}_{m,n}]=\delta_{m,0}\delta_{n,0},
\ee
imposed by the conservation of $\Tr[\rho]$.

\section{Coherent states}\label{cohsts}
It is instructive to construct the coherent states for the open harmonic oscillator. The basis $B$ can be used to define the coherent states of two oscillator,
\bea\label{cstuv}
\rho_{u,v}&=&e^{\frac{|u|^2+|v|^2}2}e^{ub^\dagger-u^*b+vb_d^\dagger-v^*b_d}\rho_0\nn
&=&e^{ub^\dagger+vb_d^\dagger}\rho_0
\eea
giving the overlap
\be
\la\la\rho^{(b)}_{m,n}|\rho_{u,v}\ra\ra=\frac{u^mv^n}{\sqrt{m!n!}}
\ee
with the basis set \eq{cdest}. These density matrices are eigenfunctions of the annihilation operators, $b\rho_{u,v}=u\rho_{u,v}$, $b_d\rho_{u,v}=v\rho_{u,v}$, are non-orthogonal,
\be
\la\la\rho_{u,v}|\rho_{u',v'}\ra\ra=e^{u^*u'+v^*v'},
\ee
provide a resolution of the identity,
\be\label{resoneuv}
\openone=\int\frac{d^2ud^2v}{\pi^2}e^{|u|^2+|v|^2}|u,v\ra\ra\la\la u,v|.
\ee
and become Hermitean for real $u$ and $v$. 

Another family of coherent states,
\bea\label{cstw}
\rho_{w_+,w_-}&=&e^{\frac{|w_+|^2+|w_-|^2}2}e^{w_+\bar a_+-w^*_+a_++w_-\bar a_--w^*_-a_-}\rho_0\nn
&=&e^{w_+\bar a_++w_-\bar a_-}\rho_0,
\eea
is defined by the help of the ladder operators. They satisfy the eigenvalue conditions $a_\pm\rho_{w_+,w_-}=w_\pm\rho_{w_+,w_-}$ and are Hermitean for $w^*_\pm=w_\mp$, $\hat\rho_w=\hat\rho_{w,w^*}=\hat\rho^\dagger_w$. The equation $\Tr[\bar a_+^{n_+}\bar a_-^{n_-}\rho_{w_+,w_-}]=0$, valid for $n_\pm\ge0$, $n_++n_->0$ can be used to prove the normalization $\Tr[\rho_{w_+,w_-}]=1$.

The matrix elements of the linear transformation between the bases $A$ and $B$, given by eqs. \eq{abtr}, can be used to establish a relation between the families \eq{cstuv} and \eq{cstw}. For this end we introduce the vectors $U=(u,v)$, $W=(w_+,w_-)$, and the matrix 
\be
T_{UW}=\begin{pmatrix}T_{\bar a_+,b^\dagger}&T^*_{\bar a_+,b^\dagger}\cr
T_{\bar a_+,b_d^\dagger}&T^*_{\bar a_+,b_d^\dagger}\end{pmatrix},
\ee
yielding
\be
\rho_{T_{UW}W}=\rho_W.
\ee
The insertion of this expression of $\rho_{u,v}$ into \eq{resoneuv} gives
\be
\openone=N\int\frac{d^2w_+d^2w_-}{\pi^2}|\rho_{w_+,w_-}\ra\ra\la\la\rho_{w_+,w_-}|
\ee
where
\be
N=2|z|^2\frac\nu{d_0+d_2}\mr{Im}\left\{\frac{\lambda_{-,-}[d_0+d_2(1+\nu\lambda_{-,+})]}{\sqrt{(d_0+d_2)^2-(d_0+d_2)2\beta\nu+d_0d_2\nu^2}}\right\}.
\ee

We record here for later use the expectation value of the canonical operators, $x$ and $p$, in a Hermitean coherent state of the ladder operators,
\bea\label{xpexpcst}
\la x\ra_w&=&\Tr[x\rho_w]=-\mr{Im}\frac{w}{2z\omega_\nu},\nn
\la p\ra_w&=&\Tr[p\rho_w]=2\mr{Re}wz.
\eea

\section{Heisenberg representation}\label{heisenbs}
The state of the system is time-independent in the Heisenberg representation hence the Schr\"odinger and the Heisenberg representations are related by a unitary similarity transformation, the time evolution operator. The time evolution is not unitary for open dynamics hence the Heisenberg representation is usually given for the full, closed  system. We are interested in the effective dynamics of the observed system therefore the Heisenberg representation should be constructed on the level of the observed, smaller system. 

The equivalence of the expectation value in the two representations, 
\be\label{scheq}
\la\psi(t)|_SA_S|\psi(t)\ra_S=\la\psi(t_i)|U^\dagger(t,t_i)A_SU(t,t_i)|\psi(t_i)\ra
=\la\psi|_HA_H|\psi\ra_H,
\ee
is the starting point to find the transformation of the state vectors and the observables of closed, unitary dynamics. The time evolution of an open system in the Schr\"odinger representation is given by the master equation \eq{master}. We assume that the equation of motion preserves the total probability. The generated time evolution,
\be
\rho_S(t)={\cal U}(t,t_i)\rho_S(t_i)
\ee
assumes the form ${\cal U}(t,t_i)=e^{(t-t_i){\cal L}}$ if the master equation is local in time. In the case of a time-dependent or non-local master equation, $\partial_t\rho(t)={\cal L}(t,t_i)\rho(t)$ with ${\cal L}(t_i,t_i)=0$, each term of the generator ${\cal L}(t,t_i)$ is assigned to the time $t$ and the solution of the equation of motion can be written in the form ${\cal U}(t,t_i)=T[e^{\int_{t_i}^tdt't{\cal L}(t',t_i)}]$ where $T$ denotes the time ordering. The inverse of the time evolution operator is well defined for finite time, ${\cal U}^{-1}(t)=e^{-t{\cal L}}$ or $\bar T[e^{-\int_{t_i}^tdt't{\cal L}(t',t_i)}]$, $\bar T$ standing for the anti-time ordering. 

The state is represented by the density matrix $\rho_H=\rho_S(t_i)$ in the Heisenberg representation and the equivalence of the expectation value of a hermitean observable,
\be\label{heeq}
\la\la\Tr|A^\dagger_S|\rho(t)\ra\ra_S=\la\la\Tr|A_S|\rho(t)\ra\ra_S
=\la\la\Tr|A_S{\cal U}(t,t_i)|\rho(t_i)\ra\ra_S
=\la\la\Tr|A_H|\rho\ra\ra_H,
\ee
where the bra $\la\la\Tr|$ is defined after \eq{expval}, can not be assured by performing a basis transformation between the two representations. The problem is formally similar to the transformation between the in-in and the in-out formalism in quantum field theory. The solution, provided by the stability of the vacuum during the time evolution in the case of field theory, $U|0\ra=|0\ra$, is realized here by the stability of the total probability, $\la\la\Tr|{\cal U}(t)|\rho\ra\ra=\la\la\Tr|\rho\ra\ra$, allowing to replace the condition \eq{heeq} by
\be
\la\la\Tr|A_S|\rho(t)\ra\ra_S=\la\la\Tr|{\cal U}^{-1}(t,t_i)A_S{\cal U}(t,t_i)|\rho(t_i)\ra\ra_S
=\la\la\Tr|A_H|\rho\ra\ra_H.
\ee
The solution,
\be
A_H(t)={\cal U}^{-1}(t,t_i)A_S{\cal U}(t,t_i),
\ee
obeys the equation of motion
\be
\partial_tA_H=[A_H{\cal L}_H],
\ee
in particular ${\cal L}_S={\cal L}_H$. 

In the case of the harmonic oscillator with local dynamics the generator \eq{generator} yields
\bea\label{aabartd}
a_{H\pm}&=&e^{t\lambda_{-,\mp}}a_\pm,\nn
\bar a_{H\pm}&=&e^{t\lambda_{+,\pm}}\bar a_\pm.
\eea
According to the basis transformation $A\to X$, given by \eq{atox}, the ladder operators can be considered as those linear superpositions of the canonical operators $x_\pm$ and $p_\pm$, building up $U$ and $U^\dagger$ in eq. \eq{densmte}, which follow an oscillatory time dependence with a normal frequency in the Heisenberg representation. For instance the expectation values \eq{xpexpcst} of the canonical operators, $x$ and $p$, in a Hermitian coherent state \eq{cstw} of the ladder operators follow the time evolution
\bea\label{xptdp}
\la x(t)\ra_w&=&-\frac{|w|}{2|z|\bar\omega_\nu}e^{-\frac\nu2t}\sin(\omega_\nu t+\phi_w-\phi_z),\nn
\la p(t)\ra_w&=&2|w||z|e^{-\frac\nu2t}\cos(\omega_\nu t-\phi_w-\phi_z),
\eea
using the notation $w=|w|e^{i\phi_w}$, $z=|z|e^{i\phi_z}$ and $\bar\omega_\nu=\sqrt{|1-\nu^2/4|}$.

The calculation of the expectation values of higher moments can be carried out by the help of Wick's theorem. The normal ordered product of the ladder operators is defined by moving $\bar a_\pm$ ($a_\pm$) to the left (right) and has vanishing expectation value in the relaxed state, $\Tr[:O:\rho_0]=0$. The contraction, $\overbrace{O}=O-:O:$, defined for the pairs of ladder operators is non-vanishing only for $\overbrace{a_\pm\bar a_\pm}=\openone$. For instance the expectation value of the product of two operators $o\cdot a=o_+a_++o_-a_-+\bar o_+\bar a_++\bar o_-\bar a_-$ is $\la(u\cdot a)(v\cdot a)\ra=u_+\bar v_++u_-\bar v_-$. In particular, the expressions \eq{x2p2} for the second moments follows in a simple manner and one can see that the disconnected components to the expectation value in a non-relaxed state receives a time-dependent damping factor $e^{-\mu t}$. Hence the energy expression of the closed oscillator approaches the asymptotic value 
\be
\hf\la p^2+x^2\ra_\infty=\frac{d_0+d_2}{2\nu}+\frac{d_2\nu}4-\frac{\beta}2,
\ee
with the dissipative relaxation time scale, $\nu$.

\section{Decoherence}\label{decs}
The decoherence denotes the suppression of the interference terms in the expectation values. This is a basis-dependent issue and is considered in the coordinate basis below. The decoherence has two different appearances: One can choose an observable $O$ and consider its expectation value in the actual state \cite{zeh,zurekd,joos}. The suppression of the contributions to the expectation value in the Schr\"odinger representation, $O_S^*(x,x_d)\rho(x,x_d,t_{obs})$, at time $t_{obs}$ with off-diagonality $x_d$ defines the instantaneous decoherence since it reflects the actual properties of the state at the observation time. Another way to look into decoherence is to single out a component of the initial state at the preparation time $t_{prep}$ with a given off-diagonality, $x_d$, and to consider its contribution to an expectation value at $t_{obs}$ \cite{dyndec}. That is more natural to realize in the Heisenberg representation. Since both the expectation value and the time evolution are linear in the Liouville space it is sufficient to retain at time $t_{prep}$ the off-diagonal component of the initial density matrix. The instantaneous and dynamical decoherence are equivalent for $t_{prep}=t_{obs}$ however the dynamical decoherence reflects the dynamical origin of decoherence, produced by the openness of the system during the time $t_{prep}<t<t_{obs}$. The characteristic time of dynamical decoherence is the dissipative time scale in simple models \cite{gas}.

The decoherence is not visible in diagonal observables hence we choose a non-diagonal $O_z=(e^{izp_d}+e^{-izp_d})/2$, given in terms of the shift operator, 
\be
e^{-izp_d}=e^{-iz(T_{p_d,a_+}a_++T_{p_d,a_-}a_-+T_{p_d,\bar a_+}\bar a_++T_{p_d,\bar a_-}\bar a_-)},
\ee
whose alternative form,
\be
e^{-izp_d}=e^{-\frac{z^2}2(T_{p_d,a_+}T_{p_d,\bar a_+}+T_{p_d,a_-}T_{p_d,\bar a_-})}e^{-iz(T_{p_d,\bar a_+}\bar a_++T_{p_d,\bar a_-}\bar a_-)}e^{-iz(T_{p_d,a_+}a_++T_{p_d,a_-}a_-)},
\ee
is used below. The filtering of the state components with off-diagonality $z$ with precision $\Delta z$ is achieved by the operator 
\be
P_{z}(t_{prep})=\hf\left[e^{-\frac1{\Delta z^2}(x_d(t_{prep})-z)^2}+e^{-\frac1{\delta z^2}(x_d(t_{prep})+z)^2}\right],
\ee
acting on the density matrix. The calculation of the desired expectation value, $\Tr[O_{z_{obs}}(t_{obs})P_{z_{prep}}(t_{prep})\rho_w]$, covering both the instantaneous and the dynamical coherence, is facilitated by the Baker-Campbell-Hausdorff formula, truncated at the level of the double commutators,
\be
e^Ae^{-B^2}=e^{-B^2}e^Ae^{-2cB+c^2},
\ee
for operators with c-number commutator, $[A,B]=c$. One finds after few straightforward steps the result
\bea\label{deckey}
\Tr[O_{z_{obs}}(t_{obs})P_{z_{prep}}(t_{prep})\rho_w]&=&e^{-\hf\la p^2\ra_0z_{obs}^2}\cos\left(\la p(t_{obs})\ra_wz_{obs}\right)\nn
&&\times\left[e^{-\frac{[z_{prep}+z_{obs}f(t_{obs}-t_{prep})]^2}{\Delta z^2}}
+e^{-\frac{[z_{prep}-z_{obs}f(t_{obs}-t_{prep})]^2}{\Delta z^2}}\right],
\eea
where $\la p^2\ra_0=\Tr[\hat p^2\rho_0]=R^2$ denotes the second moment of the momentum operator in the stationary state and
\be
f(t)=e^{-\frac\nu2t}\left(\cos\omega_\nu t-\frac\nu{2\omega_\nu}\sin\omega_\nu t\right).
\ee
The expectation value \eq{deckey} represents the weight of state components with off-diagonality $x_d=\pm z_{obs}$ at $t_{obs}$ after the components with off-diagonality $x_d\sim\pm z_{prep}$ has been filtered out at $t_{prep}$ from a coherent state, created at $t=0$.

The instantaneous decoherence of a coherent state is given by \eq{deckey} without filtering, $\Delta z=\infty$. We find a static instantaneous decoherence length scale, $\ell^2_{inst}=1/\la p^2\ra_0$. The quantity $g_{dec}^2=1/\ell^2_{inst}$ is usually interpreted as a measure of the decoherence strength, in that case it is more appropriate better to choose $g_{dec}^2=1/\ell^2_{inst}-1/2$ since the pure ground state has $\ell^2_{inst}=2$. Apart of the dominant exponential function, defining the static decoherence length scale the orthogonalization of the bra and the ket components of the state is modulated by a multiplicative factor which is oscillatory both in the off-diagonality and in the time, with characteristics length scale, given by the inverse of the expectation value of the momentum in the actual state.

The dynamical decoherence is defined by \eq{deckey} for $t_{obs}>t_{prep}$ and finite $\Delta z$. The square bracket factor indicates that the off-diagonality is changed by a multiplicative factor, $z\to z/f(t)$, during the evolution in time $t$. Thus the off-diagonality of any component of the state increases exponentially in time, the characteristic time scale being the dissipative time scale. We now choose small $\Delta z$ to make the off-diagonality at the time of the preparation  well defined and $z_{obs}=z_{prep}$ to match the instantaneous and dynamical decoherence when $t_{obs}=t_{prep}$. Then \eq{deckey} decreases in time with a double exponential \cite{dyndec}.

The surprisingly fast dynamical decoherence can be understood by considering the Euler-Lagrange equations of the Lagrangian \eq{efflagr}
\bea
\ddot x&=&-x-\nu\dot x+i(d_0x_d-d_2\ddot x_d),\nn
\ddot x_d&=&-x_d+\nu\dot x_d,
\eea
which, together with the definition of the canonical moments $p=\partial L/\partial\dot x=\dot x_d-\nu x_d$, $p_d=\partial L/\partial\dot x_d=\dot x+id_2\dot x_d-i\beta x_d$, are equivalent with the Heisenberg equations of motion of the generator \eq{genmast}. The insight, gained from the Euler-Lagrange equations, is that the time runs in opposite direction for the coordinate $x$ and for the quantum fluctuations, $x_d$. In particular, an initial off-diagonality increases in time exponentially with the dissipative time scale and suppresses any expectation value stretching over a bounded space region.

\section{Singular oscillators}\label{sings}
Despite the simplicity of the harmonic dynamics the spectrum of the open oscillator may become degenerate and the effective dynamics may display singular dependence on the environment parameters. The spectrum of an under-damped oscillator is on the unit circle of the complex frequency plane and becomes degenerate when the normal frequencies coalesce on the real or the imaginary axes, for infinitesimal system-environment interactions and at the border of the under- and over-damped oscillators, respectively.

It is well known from the degenerate perturbation expansion that weak perturbations around a degenerate matrix produce strong, non-analytic response. In a similar manner one expects singularities in the dynamics at $g=0$ and at $\nu=2$. There are actually two different eigenvalue conditions to inspect: The null-space of Neumann's equation is always degenerate, it consists of the mixtures of the stationary Liouville space vectors, the linear superpositions of projectors onto the eigenspaces of the Hamiltonian. The degeneracy is split in open systems by $\mr{Re}(\lambda)$, the finite life-time of the excitations and null-space of $\cal L$ with  $g\ne0$ is one dimensional, containing the relaxed Gaussian density matrix \eq{statdm}. The other degeneracy occurs within the four dimensional linear space of canonical operators in the construction of the ladder operators \eq{ladder}.

\subsection{Weakly open oscillator}
Both degeneracies, mentioned above, are important around $g=0$. Let us start with the closed oscillator dynamics where the generator of the time evolution of Neumann's equation, ${\cal L}^{(0)}=-i(\bar c_+c_+-\bar c_-c_-)$, displays the infinite degeneracy of the null-space. An infinitesimal system-environment interaction breaks this degeneracy and generates a singular dependence of $\rho_0$ on $g$: The parameters \eq{stdmpar} contain the ratio of the environmental parameters, $\kappa=\nu/(d_0+d_2)$, rendering the limit $g\to0$ discontinuous, $Q^2\to2\kappa$, $R^2\to1/2\kappa$ while $Q^2=1/R^2=2$ for $g=0$. 

The other degeneracy is in the space of the closed ladder operators $C$, within the subspaces $C_\pm$, span by $(\bar c_\mp,c_\pm)$, performing the shift $\Omega\to\Omega\pm i$. However an infinitesimal system-environment interaction prefers a unique basis since the ladder operators,
\bea
\bar a_\pm&\to&\pm i(\bar c_\pm-c_\mp),\nn
a_\pm&\to&\mp\frac{i}2\left[\left(\frac1\kappa+1\right)c_\pm+\left(1-\frac1\kappa\right)\bar c_\mp\right],
\eea
are singled out by the discontinuous limit of $\rho_0$ as $g\to0$. The expectation value of the energy expression of the closed harmonic oscillator, 
\be
\Tr\left[\frac{p_\pm^2+x_\pm^2}2\rho_0\right]=\frac1{2\kappa},
\ee
c.f. \eq{x2p2}, may be above or below the ground state energy of its the closed counterpart. The localisation and decoherence lengths scales are related by $\ell_{dec,\rho}=\ell_{dec,sh}=1/\ell_{loc}=\sqrt{2\kappa}$. However the practical importance of these characteristic scales is limited by the slow approach to the relaxed state, the relaxation time being $\ord{g^{-2}}$. The classical dynamics, expressed by the first moments of the canonical operators, is regular at $g=0$ and the singularity arises from the quantum fluctuations in the relaxed state.

The relaxed state of the translation invariant dynamics $\omega=0$, the quantum Brownian motion is a Gibbs operator, corresponding to an $\ord\hbar$ temperature \cite{gaussian} and it is natural to expect a similar result for the harmonic oscillator, too. Let us compare the Gibbs operator of a harmonic oscillator of mass $m_T$, frequency $\omega$ and temperature $T$,
\be
\la x^+|e^{-\frac1{k_BT}H}|x^-\ra=\frac1{\ell_T}\sqrt{\frac\hbar{2\pi\sinh\beta_\omega}}e^{\frac{-(x_f^2+x_i^2)\cosh\beta_\omega+2x_ix_f}{2\sinh\beta_\omega}},
\ee
where the dimensional coordinate is $x\ell_T$ with $\ell_T=\sqrt{\hbar/m_T\omega_\omega}$ and $\beta_\omega=\hbar\omega/k_BT$. The comparison with \eq{statdm}-\eq{stdmpar} relates the length scales,
\be
\ell_T^2=\frac{\ell^2_{cl}}{\sqrt{1+4d_0d_2}},
\ee
and gives the equation
\be
e^{\beta_\omega}=\frac{\sqrt{1+d_0d_2\kappa^2}+\kappa}{\sqrt{1+d_0d_2\kappa^2}-\kappa},
\ee
for the temperature. In the case of infinitesimal interactions the two length scales are identical, $\ell_T^2=\ell^2$, and the relaxed states within the interval $0<\kappa<1$ supports the temperature
\be
\beta_\omega=\ln\left(\frac{1+\kappa}{1-\kappa}\right).
\ee
There are non-thermal relaxed states for $1<\kappa<\kappa_0$ ias long as the upper limit, set by the positivity condition of the master equation,
\be
\kappa_0=\frac{\sqrt{2d_dd_2}}{d_0+d_2}>1,
\ee
allows.

The high temperature relaxed states are always available for small $\kappa$ however the pure ground state of the closed oscillator with $\kappa=1$ can be approached by infinitesimally weak environment interaction only if $\kappa_0\ge1$. As $\kappa$ is increased from 0 the temperature and the decoherence length drop and the localisation length increases. The pure ground state of the closed dynamics is reached at $\kappa=1$ and the relaxed state becomes more localized and recohered and ceases to be a thermal state when $\kappa>1$.

\subsection{Border between under- and over-damped oscillator}
The normal frequencies $\lambda_{\pm,+}$ and $\lambda_{\pm,-}$ become degenerate at $\nu=2$, the transition between the under- and the over-damped oscillators. Such a degeneracy takes place in the classical equation of motion, as well, whose solution is
\be
x(t)=e^{-\frac\nu2t}(e^{i\omega_\nu t}x_p+e^{-i\omega_\nu t}x_m).
\ee
The initial conditions $x(0)=x_i$, and $\dot x(0)=p_i/m$ yield 
\be\label{cssol}
x(t)=\frac{e^{-\frac\nu2t}}{2i\omega_\nu}\left[e^{i\omega_\nu t}\left(x_ii\omega_\nu+\frac{p_i}m+\frac\nu2x_i\right)+e^{-i\omega_\nu t}\left(x_ii\omega_\nu-\frac{p_i}m-\frac\nu2x_i\right)\right]
\ee
where the singular factor $1/\omega_\nu$ comes from the application of the relation $x_p-x_m=(p_i/m+x_i\nu/2)/i\omega_\nu$ in eliminating $x_p$ and $x_m$. We have the approximate regular solution,
\be
x(t)=e^{-\frac\nu2t}\left[x_i+t\left(\frac{p_i}m+\frac\nu2x_i\right)+\ord{t\omega_\nu}\right],
\ee
where the $\ord{t}$ secular solution around $\omega_\nu=0$ is reminiscent of the resonances of a forced oscillator. The solution remains regular at the degeneracy because the $1/\omega_\nu$ singularity is removed by the degeneracy of the normal modes $\lambda_{-,\pm}$. 

The degeneracy leads to singularities in the quantum case because the canonical operators are split into the ladder operators. In fact, the regular expression \eq{cssol} remains valid in the Heisenberg representation however the ladder operators are singular, c.f. eq. \eq{zfact}. since the separation of the components with different frequencies in $x$ prevents the cancellation of the singular $1/\omega_\nu$ in \eq{cssol}. The higher moments of the canonical operators induce further singularities in the quantum case. For instance we need $|w|=\ord{|2-\nu|^{1/4}}$ to keep the expectation value of the canonical operators in a coherent state regular around $\nu=2$.

\section{Summary}\label{sums}
The open Gaussian dynamics of an oscillator is investigated in this work. It involves seven parameter, the mass and the oscillator frequency parameterize the closed dynamics, Newton's friction constant represents the dissipative forces, the dynamically generated decoherence possesses two further parameter. Two additional parameters belong to time independent features, they control the asymptotic Gaussian decoherence length and characterize a gauge transformation. 

The ladder operators which shift the frequency of the density matrix by a normal frequency are constructed. The corresponding elementary excitations are described by a traceless, mixed i.e. non-factorisable component of the density matrix. The ladder operators are used write the master equation in a simple form, reminiscent of Neumann's equation and to construct coherent states.

The master equation conserves the total probability which property allows us to define the Heisenberg representation within the space of states of the observed system where the expectation value of the functions of the canonical operators is specially easy to calculate. This feature is used to locate dynamical decoherence, the system state independent part of the suppression of the interference terms in the expectation values.

While the classical damped oscillator displays regular dependence on Newton's friction coefficient the quantum fluctuations, appearing in the higher moments of the canonical operators, generate singularities for infinitesimal system-environment interactions and at the border of the under- and over-damped oscillator. 

The formalism, developed here, might be useful in different problems. One possible application is in driven quantum system \cite{sieberer}, a domain where harmonic degrees of freedom are subject of both  microscopic and macroscopic dynamics. Another, wide range of phenomemas is in weakly coupled many-body systems where the quasi-particles with a given momentum obey harmonic dynamics. In particular the creation and annihilation operators provide a simple and natural framework to follow the open dynamics of a photon mode and the coherent states yield a simple description of the laser. We hope to report on some progress in this direction soon. Another application consists of the diffusive Goldstone mode dynamics in open systems \cite{minami,hongo}.

The arguments, presented in this work raise several questions. The most pressing is the nature of the true relaxation when the initial state is far from the relaxed state. It is known that the excitations above the relaxed state die out with the dissipative time scale and the relaxation of the Gaussian states is similar process \cite{gaussian}. It remains to see how a general, non-Gaussian initial state approaches the relaxed state. 

Another question, raised by this work concerns the conditions, to be imposed on a realistic master equation. Namely the conservation of the total probability is not sufficient to guarantee the stability of the dynamics owing to the emergence of traceless components of the density matrix with exponentially growing weight. They are not present in the space of states, constructed from the relaxed Gaussian state by the help of the ladder operators but can be generated if there is a relaxation in the Gaussian state. How to recognize the existence of such states and to make sure that they are excluded? 

An open system can be turned into the testing ground of thermodynamics and the formalism, introduced here, may help to construct simple models to demonstrate the emergence of thermodynamics within closed quantum systems \cite{gemmer,kosloff}. Further possible avenues are the clarification of some unusual features of strongly damped systems \cite{haake}, of the equipartition theorem \cite{bialas} and of the entropy production \cite{alipour,pucci}.

Yet another question is raised by the family of relaxed states of harmonic systems, governed by a local master equation. The translation invariant open dynamics of a free particle with infinitesimal system-environment coupling relaxes to a thermal equilibrium state. However the harmonic oscillator with infinitesimal system-environment coupling can relax to non-thermal states, as well. In general, the relaxed state can be more or less localized and more or less decohered than the pure ground state. It would be interesting to identify the dynamical mechanism which stabilizes these unusual states.

Finally, the present results can be the starting point to develop the Heisenberg and the interaction representations in open quantum field theories. However to be useful, these constructions must cover non-local interactions beyond the expansion in the time derivative.


\begin{thebibliography}{99}
\bibitem{agarwal} G. S. Agarwal, \journal{Phys. Rev.}{A4}{739}{1971}.
\bibitem{dattagupta} S. Dattagupta, \journal{Phys. Rev.}{A30}{1525}{1984}.
\bibitem{gardinerc} C. W. Gardiner, M. J. Collett, \journal{Phys. Rev.}{A31}{3761}{1985}.
\bibitem{ford} G. W. Ford, M. Kac, \journal{J. Stat. Phys.}{46}{803}{1987}.
\bibitem{walls} D. F. Walls, G. J. Milburn, \journal{Phys. Rev.}{A31}{2403}{1985}.
\bibitem{savage} C. M. Savage, D. F. Walls, \journal{Phys. Rev.}{A32}{2316}{1985}.
\bibitem{caldeira}  A. O. Caldeira, A. J. Leggett, \journal{Physica}{121A}{587}{1983}; \journal{Ann. Phys. (N.Y.)}{149}{374}{1983}; \journal{Phys. Rev.}{A31}{1059}{1985}.
\bibitem{haake} F. Haake, R. Reibold, \journal{Phys. Rev.}{A32}{2462}{1985}.
\bibitem{riseborough} P. S. Riseborough, P. H\"anggi, U. Weiss, \journal{Phys. Rev.}{A31}{471}{1985}.
\bibitem{dekker} H. Dekker, \journal{Phys. Rep.}{80}{1}{1981}.
\bibitem{grabert} H. Grabert, P. Schramm, G. L. Ingold, \journal{Phys. Rep.}{168}{115}{1988}.
\bibitem{boyankovsky} D. Boyankovsky, D. Jasnow, \journal{Phys. rev.}{A96}{062108}{2017}.
\bibitem{nakajima} S. Nakajima, \journal{Progr. Teor. Phys.}{20}{948}{1958}.
\bibitem{zwanzig} R. Zwanzig, \journal{J. Chem. Phys.}{33}{1338}{1960}.
\bibitem{esposito} M. Esposito, \journal{Phys. Rev.}{E68}{066112}{2003}.
\bibitem{budini} A. A. Budini, \journal{Phys. Rev.}{E72}{056106}{2005}.
\bibitem{breuerpr} H. P. Breuer, \journal{Phys. Rev.}{A75}{022103}{2007}.
\bibitem{vacchini} B. Vacchini, \journal{J. Phys.}{B45}{154007}{2012}.
\bibitem{grabert84} H. Grabert, U. Weiss, P. Talkner, \journal{Z. Phys.}{B55}{87}{1984}.
\bibitem{karrlein} R. Karrlein, H. Grabert, \journal{Phys. Rev.}{E55}{153}{1977}.
\bibitem{gardiner} C. W. Gardiner, P. Zoller, {\em Quantum Noise}, 2. enl. ed. Springer, Berlin (2000).
\bibitem{breuer} H.-P. Breuer, F. Petrucciona, {\em The Theory of Open Quantum Systems}, Oxford University Press, Oxford (2002).
\bibitem{alicki} R. Alicki, M. Fannes, {\em Quantum dynamical systems}, Oxford University Press, Oxford (2001).
\bibitem{banerjee} S. Banerjee, {\em Open Quantum Systems}, Springer, Hindustan Book Agency, New Delhi (2018).
\bibitem{alf} J. Polonyi, \journal{Int. J. Mod. Phys.}{34}{1950077}{2019}.
\bibitem{ostrogadsky} M. Ostrogradsky, \journal{Mem. Ac. St. Petersbourg}{VI 4}{385}{1850}; P. Woodard, {\em Ostrogradsky's theorem on Hamiltonian instability}, \journal{Scholarpedia}{10}{32243}{2015}, DOI: 10.4249/scholarpedia.32243, arXiv:1506.02210.
\bibitem{tay} B. A. Tay, T. Petrosky, \journal{J. Math. Phys.}{49}{113301}{2008}.
\bibitem{prosen} T. Prosen, \journal{New J. of Phys.}{10}{043026}{2008}.
\bibitem{honda} D. Honda, H. Nakazoto, M. Yoshida, \journal{J. Math. Phys.}{51}{072107}{2010}.
\bibitem{schw} J. Schwinger, \journal{J. Math. Phys.}{2}{407}{1961}.
\bibitem{keldysh} L. V. Keldysh, \journal{Zh. Eksp. Teor. Fiz.}{47}{1515}{1964}
(\journal{Sov. Phys. JETP}{20}{1018}{1965}).
\bibitem{hu9293} B. L. Hu, J. P. Paz, Y. Zhang, \journal{Phys. Rev.}{D45}{2843}{1992}; {\em ibid} \journal{Phys. Rev.}{D47}{1576}{1993}.
\bibitem{stienspring} W. F. Stinespring, \journal{Proc. Am. Math. Soc.}{6}{211}{1955}.
\bibitem{kraus} K. Kraus, \journal{Ann. Phys.}{64}{311}{1971}.
\bibitem{gorini} V. Gorini, A. Kossakowski, E. C. G. Sudarshan, \journal{J. Math. Phys.}{17}{821}{1976}.
\bibitem{lindblad} G. Lindblad, \journal{Comm. Math. Phys.}{48}{119}{1976}.
\bibitem{sandulescu} A. Sandulescu, H. Scutaru, \journal{Ann. Phys. (N. Y.)}{173}{277}{1987}.
\bibitem{isar} A. Isar, A. Sandulescu, H. Scutaru, E. Stefanscu, W. Scheid, \journal{Int. J. Mod. Phys.}{E3}{635}{1994}.
\bibitem{schulman} L. S. Schulman, {\em Techniques and Application of Path Integration}, J. Wiley \& Sons, New York, 1981.
\bibitem{free} J. Polonyi. I. Rachid, {\em Equilibrium particle states in weakly open dynamics}, arXiv:1904.06338 .
\bibitem{dyndec} J. Polonyi, \journal{J. of Phys.}{A51}{145302}{2018}.
\bibitem{zeh} H. D. Zeh, \journal{Found. Phys.}{1}{69}{1970}.
\bibitem{zurekd} W. H. Zurek, \journal{Phys. Rev.}{D24}{1516}{1981}.
\bibitem{joos} E. Joos, H. D. Zeh, \journal{Z. Phys.}{B59}{223}{1985}.
\bibitem{gas} J. Polonyi, \journal{Phys .Rev.}{A92}{042111}{2015}.
\bibitem{masterl} R. Zwanzig, \journal{Physica}{30}{1109}{1964}.
\bibitem{mastertf} H. Umezawa, H. Matsumoto, M. Tachiki, {\em Thermo Field Dynamicsand Condensed States} North-Holland, Amsterdam (1982); N. P. Landsman, Ch. G. Weert, \journal{Phys. Rep.}{145}{141}{1987}.
\bibitem{sieberer} L. M. Sieberer, M. Buchhold, S. Diehl, \journal{Rep. Progr. Phys.}{79}{096001}{2016}.
\bibitem{minami} Y. Minami, Y. Hidaka, \journal{Phys. Rev. Lett.}{97}{012130}{2018}.
\bibitem{hongo} M. Hongo, S. Kim, T. Noumi, A. Ota, \journal{JHEP}{02}{131}{2019}.
\bibitem{gaussian} J. Polonyi, {\em A Gaussian density matrix under decoherence and friction}, arXiv:1510.03212.
\bibitem{gemmer} J. Gemmer, M. Michel, G. Mahler, {\em Quantum Thermodynamics}, Springer, Berlin (2009).
\bibitem{kosloff} R. Kosloff, \journal{Entropy}{15}{2100}{2013}.
\bibitem{bialas} P. Bialas, J. Spechowicz, L. Lucaka, \journal{J. Phys.}{A52}{15LT01}{2019}; \journal{Sci. Rep.}{8}{16080}{2018}.
\bibitem{alipour} S. Alipour et. al. \journal{Sci. Rep.}{6}{35568}{2016}.
\bibitem{pucci} L. Pucci, M. Esposito, L. Peliti, \journal{J. Stat. Mech.}{}{P04005}{2013}.
\end{thebibliography}
\end{document}